\documentclass[10pt,aps,prd,twocolumn,fleqn,floatfix,nofootinbib]{revtex4-2}

\usepackage[section]{placeins}
\usepackage{amsmath,amssymb,amsfonts}
\usepackage{graphicx}
\usepackage[usenames,dvipsnames]{xcolor}
\usepackage{mathtools}        
\usepackage{bm}                 
\usepackage{microtype}          
\usepackage[breaklinks=true,colorlinks=true,
            linkcolor=MidnightBlue,citecolor=MidnightBlue,urlcolor=MidnightBlue]{hyperref}

\allowdisplaybreaks[1]

\graphicspath{{./figs/}{./}}
\newcommand{\SYM}{\mathcal N=4~\mathrm{SYM}}
\newcommand{\SoverS}{\mathcal S/\mathcal S_0}
\newcommand{\PoverP}{p/p_0}
\newcommand{\EoverE}{\varepsilon/\varepsilon_0}
\newcommand{\Lam}{\lambda}
\newcommand{\Li}{\log\!\frac{\Lam}{\pi^2}}

\begin{document}

\title{Constrained Pad\'e Ensembles for Thermal \texorpdfstring{$\mathcal N=4$}{N=4} SYM:
Quantified Uncertainties and Next-Order Predictions}

\author{Ubaid Tantary}
\email{utantary@pmu.edu.sa}
\affiliation{Department of Mathematics and Natural Sciences, Prince Mohammad Bin Fahd University, Al Khobar 31952, Saudi Arabia}

\date{\today}

\begin{abstract}
We quantify the transition between weak and strong coupling in thermal ${\cal N}=4$ supersymmetric Yang–Mills (SYM) theory in four space-time dimensions by constructing an \emph{admissible ensemble} of log-aware Pad\'e approximants that incorporate the weak- and strong-coupling expansions through $\mathcal O(\Lam^2)$ and $\mathcal O(\Lam^{-3/2})$ ($\Lam$ is the 't~Hooft coupling), including the non-analytic $\Lam^{3/2}$ and $\Lam^{2}\log\Lam$ terms. This replaces single-curve estimates with a reproducible uncertainty band and a well-defined central curve across the intermediate regime. The framework is \emph{predictive}, setting testable benchmarks for forthcoming perturbative and holographic calculations.
\end{abstract}

\maketitle

\section{Introduction}
The thermodynamics of ${\cal N}=4$ supersymmetric Yang--Mills theory in four dimensions ($\SYM$) is a useful benchmark for interpolation across coupling. Conformality implies that the Stefan-Boltzmann-normalized ratios of pressure, energy density, and entropy density coincide,
\begin{equation}
\begin{aligned}
&\qquad\qquad \PoverP=\EoverE=\SoverS=:f(\Lam),\\
&\qquad\qquad \varepsilon-3p=0,\qquad c_s^2=\tfrac13,
\end{aligned}
\end{equation}
so a single function exhausts equilibrium thermodynamics with $\mathcal S_0=\frac{2\pi^2}{3}\,d_A\,T^3$, $\mathcal F_0=-p_0=-\frac{\pi^2}{6}\,d_A\,T^4$, and with $d_A=N_c^2-1$ being the dimension of the adjoint representation. We work with the ratios $f(\Lam)=\SoverS=\PoverP=\EoverE$ throughout. On the weak side, the $\mathcal O(\Lam^2)$ expansion with its exact nonanalytic structure was obtained by direct/HTL resummation and independently reproduced via an EFT construction~\cite{Du:2021jai,Andersen:2021kld}; on the strong side, the large-$\Lam$ expansion at large $N_c$ follows from AdS/CFT~\cite{Maldacena:1997re,Gubser:1998bc}.

Previous Pad\'e studies provided \emph{point estimates} without uncertainty bars, leaving the robustness of crossover predictions unclear. Earlier work used single near-diagonal Pad\'es, which are sensitive to matching choices and to the weak-side logarithm. Precisely in the intermediate regime where weak-coupling ceases to converge ($\Lam\gtrsim 1$) but strong-coupling corrections are still sizable ($\Lam\lesssim 10$), predictions become particularly dependent on interpolation choices. This window overlaps the phenomenologically relevant range for hot, strongly interacting matter.

Our contribution is to replace single-curve Pad\'es with an admissible \emph{ensemble} and to report a \emph{model band}. We develop two independent \emph{log-aware} routes: (i) a Hermite-Pad\'e (HP) interpolant that matches the generalized two-point Pad\'e of Ref.~\cite{Du:2021jai}, including the exact $\Lam^{2}\log\Lam$ term and the $4/3$ factors enforcing $f\to 3/4$; and (ii) a log-subtracted two-point Pad\'e (LSTP) that removes the known $\Lam^{2}\log\Lam$ term before fitting a rational approximant to the remainder. Both satisfy standard \emph{admissibility constraints}: no poles on $\Lam>0$, bounded within $0.75\le f\le 1$, and monotone in $\log\Lam$. The surviving ensemble quantifies interpolation uncertainty with a reproducible band and a well-defined central curve, enabling next-order predictions. The result is a quantitatively defensible crossover with transparent model dependence.

Motivated by suggestions raised in private correspondence (2021) after our $\mathcal O(\Lam^2)$ work~\cite{Du:2021jai}, we asked whether a Pad\'e construction could \emph{predict} the next strong-coupling coefficient using only the weak-coupling expansion through $\mathcal{O}(\Lam^2)$ and the leading holographic correction $\mathcal{O}(\Lam^{-3/2})$. While we initially expected that one more weak-side order would be necessary, the admissible \emph{ensemble} developed here (log-aware, pole-free, and bounded) shows that such predictions are feasible: we extract $A_{5/2}$ and constrain $\widehat S_{3}(\Lam_\ast)$ with quantified model uncertainties.

Both our ensemble approach and Ref.~\cite{Muller:2025} interpolate between the $\mathcal{O}(\Lam^2)$ weak-coupling expansion~\cite{Du:2021jai,Andersen:2021kld} and the $\mathcal{O}(\Lam^{-3/2})$ strong-coupling expansion~\cite{Gubser:1998bc}, and both use a curvature-based diagnostic to identify a pseudocritical coupling $\Lam_c$. The methodological difference is that we replace a single interpolant with an \emph{admissible ensemble}, yielding a reproducible uncertainty band (and central curve) rather than a single point estimate.

\section{Weak- and strong-coupling expansions }\label{sec:inputs}
We use the 't Hooft coupling $\Lam=g^2N_c$.  At weak coupling the Stefan-Boltzmann-normalized entropy ratio is
\begin{equation}
\label{eq:weak}
\begin{aligned}
\quad f(\Lam)
&= 1 - \frac{3}{2\pi^2}\,\Lam
   + \frac{3+\sqrt{2}}{\pi^3}\,\Lam^{3/2}+ \frac{3}{2\pi^4}\,\Lam^2\,\Li
  \\
&\quad  + A_{20}\,\Big(\tfrac{\Lam}{\pi^2}\Big)^{\!2}
   + \mathcal O(\Lam^{5/2})\,,
\end{aligned}
\end{equation}

where
\[
A_{20}
= -\frac{21}{8} \;-\; \frac{9\sqrt{2}}{8}
  \;+\; \frac{3}{2}\,\gamma_E
  \;+\; \frac{3}{2}\,\frac{\zeta'(-1)}{\zeta(-1)}
  \;-\; \frac{25}{8}\,\log 2\,,
\]

and the logarithmic coefficient multiplying $\Lam^2\log(\Lam/\pi^2)$ is exactly $A_{2\log}=\tfrac{3}{2\pi^4}$.
The constants $A_{20}$ and $A_{2\log}$ were obtained and cross-checked by direct resummation~\cite{Du:2021jai} and by an EFT reconstruction~\cite{Andersen:2021kld}.

At strong coupling (large $N_c$)~\cite{Gubser:1998bc},
\begin{equation}
\label{eq:strong}
\qquad f(\Lam)=\frac{3}{4}\Big[1+\frac{15}{8}\zeta(3)\,\Lam^{-3/2}+\mathcal O(\Lam^{-3})\Big],
\end{equation}
with no $\Lam^{-1/2}$ or $\Lam^{-1}$ terms.

\section{Log–aware conformal Pad\'e methodology}\label{sec:method}
Let $y=\sqrt{\Lam}$ and map the positive axis via
\begin{equation}
\label{eq:map}
\qquad z=\frac{y}{1+\alpha y+\beta y^2},\qquad \alpha>0,\ \beta\ge0,
\end{equation}
which includes one-parameter maps at $\beta=0$. Rational/conformal mappings are routinely
used to suppress spurious poles and compactify semi–infinite domains~\cite{BoydSpectral,BakerGravesMorris}.
We use two complementary routes.

\subsection{Route A: log–subtracted two–point Pad\'e (LSTP)}
Define
\begin{equation}
\begin{aligned}
\qquad g(\Lam)&=f(\Lam)-\frac{3}{2\pi^4}\Lam^2\Li\,\chi(\Lam;\Lambda_0,p),\\
&\qquad\chi(\Lam;\Lambda_0,p)=\frac{1}{1+(\Lam/\Lambda_0)^p},
\end{aligned}
\end{equation}
with integer $p\ge1$ so the subtraction is exact at small $\Lam$ but dies off at large $\Lam$.
Subtracting the logarithm and rationally approximating the residual is standard in series
analysis (Dlog/Pad\'e-type preprocessing)~\cite{Guttmann1989,BakerGravesMorris}. We employ a
smooth cutoff $\chi$ to decouple the weak-side logarithm from the strong-side constraints~\cite{BoydSpectral}.
We then approximate
\begin{equation}
\label{eq:LSTP}
\qquad \qquad \qquad g(\Lam)\approx \frac{P_m(z)}{Q_n(z)},
\end{equation}
where
\begin{equation}
\qquad P_m(z)=\sum_{k=0}^m p_k z^k,\quad Q_n(z)=1+\sum_{k=1}^n q_k z^k,
\end{equation}
and set
\begin{equation}
\qquad f(\Lam)\approx \frac{P_m(z)}{Q_n(z)}+\frac{3}{2\pi^4}\Lam^2\Li\ \chi(\Lam;\Lambda_0,p)\, .
\end{equation}
Coefficients are fixed by expanding about $\Lam\to0$ and $\Lam\to\infty$ and matching
Eqs.~\eqref{eq:weak}-\eqref{eq:strong}. We scan near-diagonal orders $[m/n]=[4/4]$ and mapping
parameters $(\alpha,\beta)$, with $(\Lambda_0,p)$ controlling the cutoff.

\paragraph{Choice of $p$ and strong-coupling artifact.}
The cutoff $\chi(\Lam)=1/(1+(\Lam/\Lambda_0)^p)$ leaves the weak-coupling
subtraction exact through $\mathcal O(\Lam^2\Li)$ for any $p\ge1$
(deviations first appear at $\mathcal O(\Lam^{2+p}\Li)$).
At large $\Lam$, however, the cutoff tail $\chi\sim(\Lambda_0/\Lam)^p$
induces a leakage
\begin{equation}
\label{eq:LSTP_leak}
\qquad\Delta f_{\mathrm{leak}}=\frac{3}{2\pi^4}\,\Lambda_0^{p}\,\Lam^{2-p}\,\Li\,.
\end{equation}
For $p=1$ or $2$ this term does not decay ($\sim\Lam\,\Li$ or $\sim\Li$,
respectively); $p=3$ is the \emph{minimal integer} for which the leakage
decays at large $\Lam$.  Setting $p=3$ in Eq.~\eqref{eq:LSTP_leak} gives
\begin{equation}
\label{eq:leak_p3}
|\Delta f_{\mathrm{leak}}|\;\le\;\frac{3}{2\pi^4}\,\frac{\Lambda_0^{3}}{\Lam}\,\Big|\Li\Big|
\;\lesssim\;
\begin{cases}
1.7\times10^{-2} & (\Lam=5),\\[2pt]
5\times10^{-3}   & (\Lam\ge10),
\end{cases}
\end{equation}
for $\Lambda_0\le2$, well within the admissible-band width.
Increasing $p$ suppresses the leakage faster
(e.g.\ $\Lam^{-2}\Li$ for $p=4$) but makes the cutoff steeper
near $\Lambda_0$.
Repeating the full admissible scan with $p=4$ yields an envelope
that differs by $\sup|\Delta f|\approx0.010$ from the $p=3$ band
($\lesssim 1.1\%$ in $\mathcal S/\mathcal S_0$).
For $p=5$ the sharper cutoff narrows the band (fewer candidates
pass the admissibility filters), but the central tendency is consistent.
We adopt $p=3$ as the minimal well-motivated choice and make the resulting
artifact explicit rather than hidden.

\subsection{Route B: two-point (Hermite-Pad\'e) rational approximant (HP)}

We use the generalized two-point Pad\'e form introduced in Appendix~G of
Ref.~\cite{Du:2021jai} (see also Appendix~C of Ref.~\cite{Andersen:2021kld}):
\begin{equation}
\label{eq:HP}
f(\Lam)=\frac{1+a\Lam^{1/2}+b(\Lam)\Lam+c\,\Lam^{3/2}+d\,\Lam^2+e(\Lam)\Lam^{5/2}}
{1+a\Lam^{1/2}+\bar b(\Lam)\Lam+\tfrac{4}{3}c\,\Lam^{3/2}+\tfrac{4}{3}d\,\Lam^2+\tfrac{4}{3}e(\Lam)\Lam^{5/2}}.
\end{equation}
The factor $4/3$ in the denominator enforces $f\to3/4$ at large $\Lam$ while
eliminating $\Lam^{-1/2}$ and $\Lam^{-1}$ terms in the strong-coupling expansion.
The shifted coefficient $\bar b(\Lam)=b(\Lam)+\tfrac{3}{2\pi^2}$ ensures the
$\mathcal O(\Lam)$ term matches Eq.~\eqref{eq:weak}.
Constants $a,c,d$ and the non-log parts of $b,e$ are uniquely fixed by matching
Eqs.~\eqref{eq:weak} and~\eqref{eq:strong}; this is a two-point (Hermite) Pad\'e
approximant in the sense of Refs.~\cite{BakerGravesMorris,Chisholm1973}.
The coefficients are
\begin{align}
\label{eq:HP_coeffs}
a &= \frac{4\pi^2}{135\,\zeta(3)}+\frac{2(3{+}\sqrt{2})}{3\pi}
\approx 1.1800\,,\nonumber\\[4pt]
b(\Lam) &= b_0 + \frac{1}{\pi^2}\log\!\frac{\Lam}{\pi^2}\,,\displaybreak[3]\nonumber\\[4pt]
b_0 &= \frac{16\pi\bigl[45(3{+}\sqrt{2})\zeta(3)+\pi^3\bigr]}{18225\,\zeta(3)^2}\nonumber\\*
&\quad+\frac{36\bigl[\frac{\zeta'(-1)}{\zeta(-1)}+\gamma_E\bigr]+69\sqrt{2}+59-75\log2}{36\pi^2}\nonumber\\*
&\approx 1.0689\,,\nonumber\\[4pt]
\bar b(\Lam) &= b(\Lam)+\frac{3}{2\pi^2}\,,\nonumber\\[4pt]
c &= \frac{2}{15\,\zeta(3)}\approx 0.1109\,,\nonumber\\[4pt]
d &= \frac{180(3{+}\sqrt{2})\zeta(3)+8\pi^3}{2025\,\pi\,\zeta(3)^2}\approx 0.1309\,,\nonumber\\[4pt]
e(\Lam) &= \frac{2\,b(\Lam)}{15\,\zeta(3)}-\frac{3}{5\pi^2\zeta(3)}\,.
\end{align}
The $\log(\Lam/\pi^2)$ in $b(\Lam)$ reproduces the exact $\Lam^2\log\Lam$ coefficient.
The coefficient $e(\Lam)$ inherits a logarithmic part from $b(\Lam)$; this
is not fixed by the current perturbative expansion and is a consequence of the
rational structure of the ansatz.

\paragraph{Strong-coupling structure of the HP.}
At large $\Lam$, all three matched strong-coupling constraints are
reproduced \emph{exactly} by Eq.~\eqref{eq:HP}, not merely asymptotically:
(i) $f\to3/4$, (ii) the coefficients of $\Lam^{-1/2}$ and $\Lam^{-1}$
vanish identically for any value of $\log\Lam$ (enforced by the $4/3$
factors in the denominator), and (iii) the coefficient of $\Lam^{-3/2}$
equals $S_{3/2}=\tfrac{15}{8}\zeta(3)$ exactly.
The last identity follows because the log-dependent parts of $b(\Lam)$
and $e(\Lam)$ cancel in the $\Lam^{-3/2}$ coefficient:
defining $e_0=2b_0/[15\zeta(3)]-3/[5\pi^2\zeta(3)]$ (the non-log part of $e$),
one verifies $\pi^{2}b_0-\tfrac{9}{2}=\tfrac{15}{2}\pi^{2}\zeta(3)\,e_0$
analytically.
Moreover, the coefficients of $\Lam^{-2}$ and $\Lam^{-5/2}$ vanish
exactly---one verifies algebraically that $2a=15d\,\zeta(3)$ and
$15c\,\zeta(3)=2$---so the next nonvanishing strong-coupling
contribution appears at $\mathcal O(\Lam^{-3})$, consistent with the
large-$N_c$ structure in which the expansion proceeds in powers of
$\Lam^{-3/2}$~\cite{Gubser:1998bc}.
Log artifacts enter only at this order: the $\Lam^{-3}$ coefficient
(the HP prediction for $S_3$) carries an $\mathcal O(1/\Li)$ correction
from the rational structure.
Crucially, no multiplicative $\Lam^{-n}\Li$ artifact is generated
through the orders we have checked; the contamination is $1/\log$-suppressed and decays
monotonically for $\Lam\gg1$, so it does not spoil the admissibility
filters.

\subsection{Admissibility filters and band}\label{sec:filters}
Candidates must satisfy, on $\Lam\in[\Lam_{\min},\Lam_{\max}]$ evaluated on a logarithmic grid:
\begin{enumerate}
\item \textit{Bounds:} $0.75\le f(\Lam)\le 1$.
\item \textit{Monotonicity in log space:} $\tfrac{d f}{d(\log\Lam)}\le 0$.
\item \textit{Pole exclusion:} compute all roots of $Q_n(z)$ and (for HP) the full denominator, map them to the $\Lam$ plane using~\eqref{eq:map}, and reject any pole on the positive real axis. We also reject near-canceling Froissart doublets (root-pole pairs whose separation is numerically indistinguishable on the grid).
\end{enumerate}
The surviving set $\{f_i\}$ defines the \emph{admissible band} $[f_{\min}(\Lam),f_{\max}(\Lam)]$ and a central curve
\begin{equation}
\label{eq:central}
f_{\mathrm{cent}}=\arg\min_{i}\ \int_{\log\Lam_{\min}}^{\log\Lam_{\max}}
\left(\frac{d^{2} f_i}{d(\log\Lam)^2}\right)^{\!2}\, d(\log\Lam).
\end{equation}
We define the crossover as the inflection in log space,
\begin{equation}
\label{eq:lc}
\qquad\qquad\qquad\frac{d^{2} f}{d(\log\Lam)^2}\Big|_{\Lam=\Lam_c}=0,
\end{equation}
choosing the zero nearest to the largest curvature peak if multiple inflections exist.
\section{Equilibrium thermodynamics: admissible band, crossover, and higher–order predictions}
\label{sec:eq}
\subsection{Ensemble and central solution}
The HP generalized Pad\'e passes all filters and minimizes the curvature functional in Eq.~\eqref{eq:central}; we take it as the central curve. 
For the LSTP route with near-diagonal $[m/n]=[4/4]$ and $p=3$, the scan yields \emph{nine} admissible survivors when $\beta=0$, at
\begin{equation}
\label{eq:LSTP_grid}
\qquad\alpha\in\{0.5,1.0,2.0\},\qquad \Lambda_0\in\{0.5,1.0,2.0\},
\end{equation}
with no survivors at $\Lambda_0=4.0$. Cases with $\beta>0$ typically develop positive-axis poles or violate the bounds and are rejected.
For $\beta=0$ the M\"obius map $z=y/(1+\alpha y)$ reparametrizes the
same $[4/4]$ rational function class, so the nine parameter sets
yield \emph{three distinct curves} (one per $\Lambda_0$); the
$\alpha$-independence serves as an internal consistency check
(cf.\ Tables~\ref{tab:summary} and~\ref{tab:polemargins}).

\paragraph{\textbf{Crossover scales.}}
For the central HP curve,
\begin{equation}
\label{eq:lambda_c_center}
\qquad \Lam_c^{\mathrm{center}} \simeq 3.5223,
\qquad
f\!\left(\Lam_c^{\mathrm{center}}\right) \simeq 0.8539.
\end{equation}
Across the admissible ensemble,
\begin{equation}
\label{eq:lambda_c_band}
\qquad\Lam_c \in [2.9520,\,6.7321],
\qquad
f(\Lam_c)\in[0.8345,\,0.8609].
\end{equation}

\paragraph{\textbf{Pole safety.}}
All LSTP survivors are free of poles on $\Lam>0$ (mapping in Eq.~\eqref{eq:map}). 
The nearest mapped poles lie well away from the positive axis; the minimal imaginary part satisfies 
$\operatorname{Im}\Lam \ge 7.38$ across survivors.

\paragraph{\textbf{Physical interpretation.}}
The central crossover, defined by the inflection in $\log\Lam$ [Eq.~\eqref{eq:lc}], occurs at $\Lam_c\simeq 3.52$, where $f(\Lam_c)\simeq 0.854$---i.e.\ the entropy density is $\sim 85\%$ of the ideal value, indicating substantial interaction effects already at moderate coupling. 
The admissible \emph{range} $\Lam_c\in[2.95,\,6.73]$ is not statistical noise; it reflects genuine model dependence given the present expansion (no $\mathcal O(\Lam^{5/2})$ term on the weak side and no $\mathcal O(\Lam^{-3})$ correction on the strong side). 
Interpreting any single-curve Pad\'e without an uncertainty band would therefore overstate precision precisely in this intermediate-coupling regime; the admissible ensemble renders this uncertainty explicit and reproducible.

\begin{figure}[htb]
  \centering
  \begin{minipage}[t]{0.5\linewidth}
    \centering
    \includegraphics[width=\linewidth]{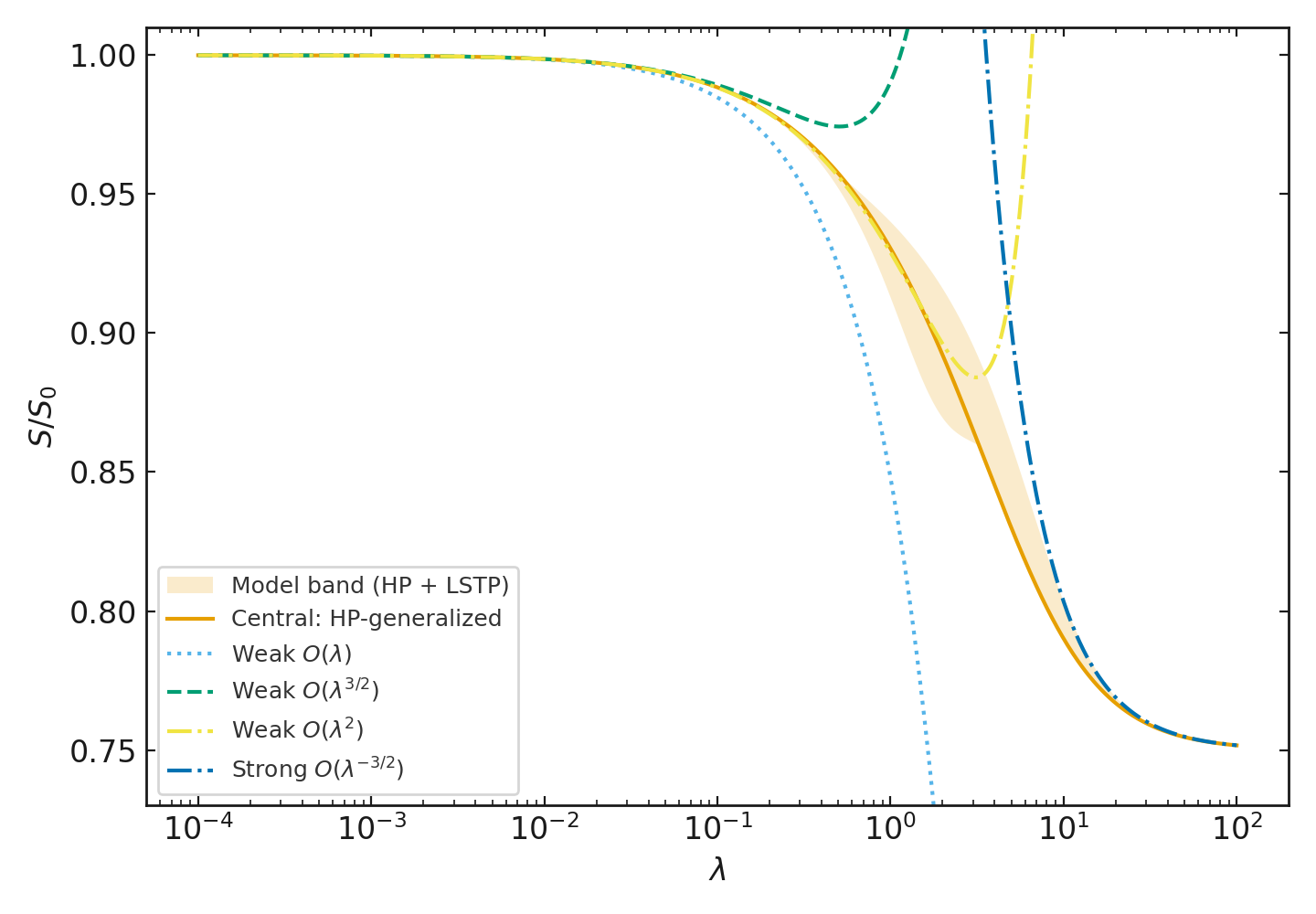}
  \end{minipage}\hfill
  \begin{minipage}[t]{0.5\linewidth}
    \centering
    \includegraphics[width=\linewidth]{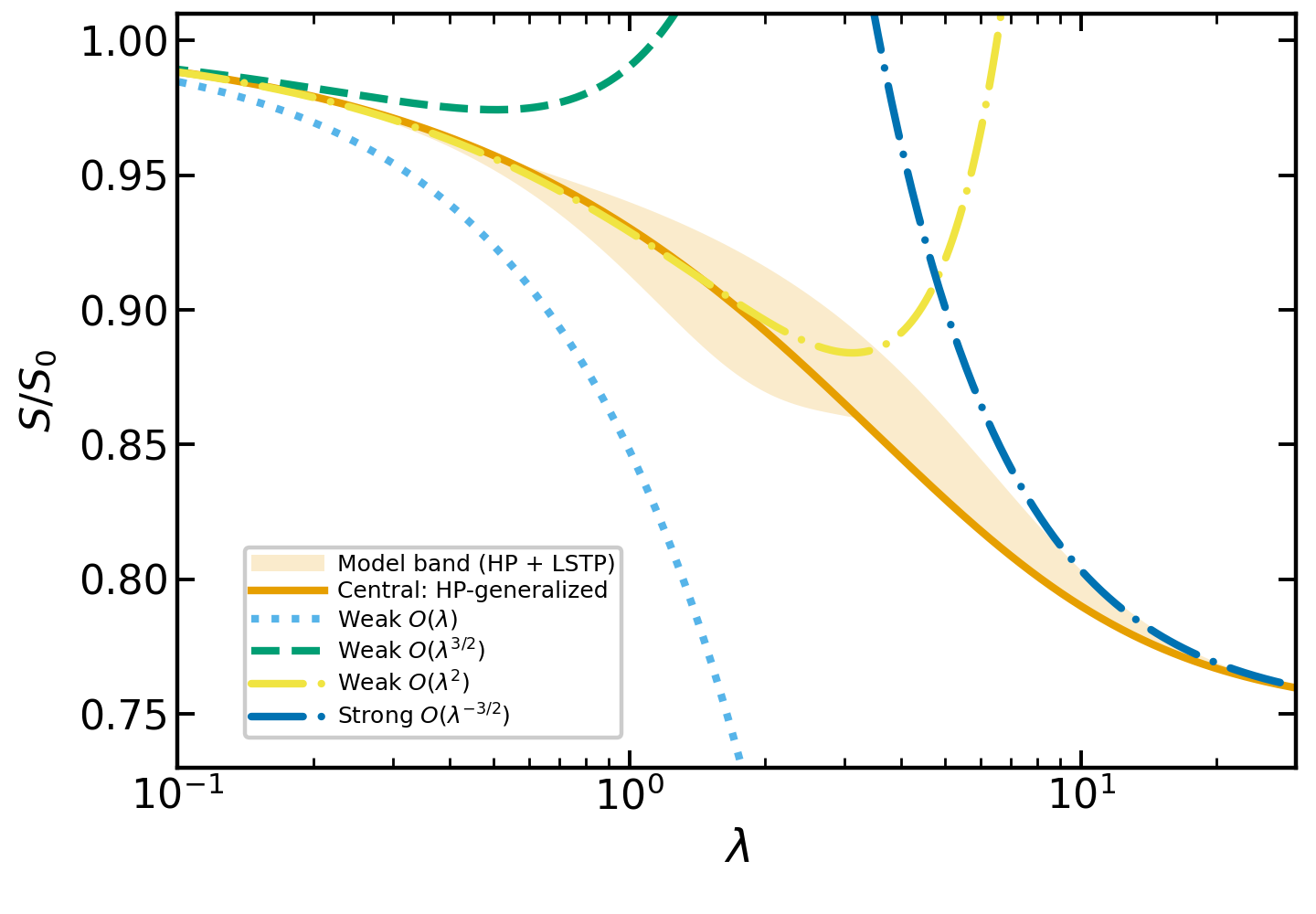}
  \end{minipage}

  \caption{Admissible Pad\'e band for $f(\Lam)=\SoverS$ in $\SYM$.
  \textbf{Left:} full-range view.
  \textbf{Right:} zoomed view emphasizing the intermediate
region $0.2\lesssim\Lam\lesssim 10$ where most of the variation occurs.
  Shaded: band; solid: central curve. Also shown are the weak truncations
  $\mathcal O(\Lam)$, $\mathcal O(\Lam^{3/2})$, $\mathcal O(\Lam^{2})$
  (including the exact $\Lam^{2}\log\Lam$ term) and the strong truncation
  $\mathcal O(\Lam^{-3/2})$.}
  \label{fig:band}
\end{figure}

\begin{figure}[htb]
  \centering
  \begin{minipage}[t]{0.5\linewidth}
    \centering
    \includegraphics[width=\linewidth]{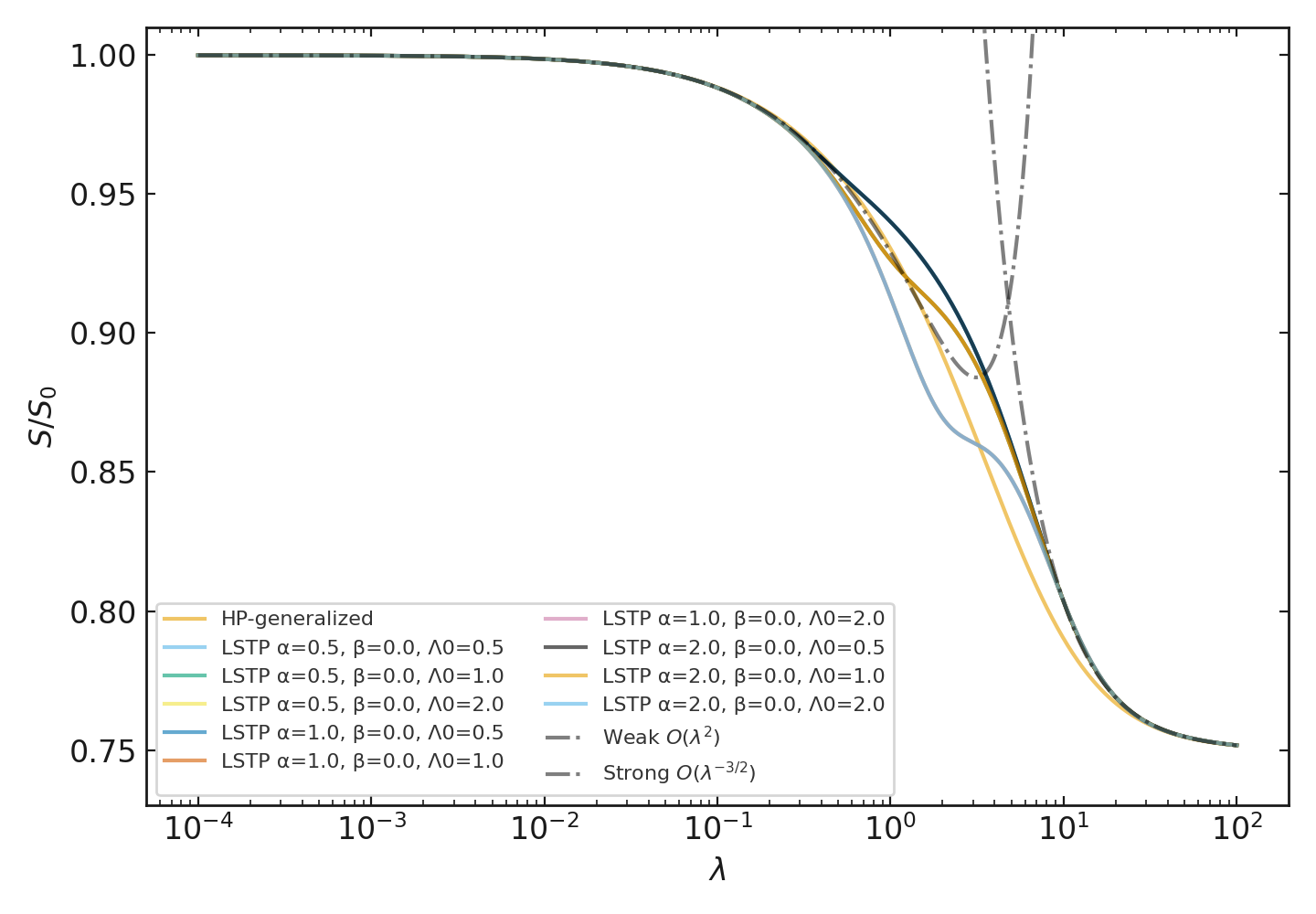}
  \end{minipage}\hfill
  \begin{minipage}[t]{0.49\linewidth}
    \centering
    \includegraphics[width=\linewidth]{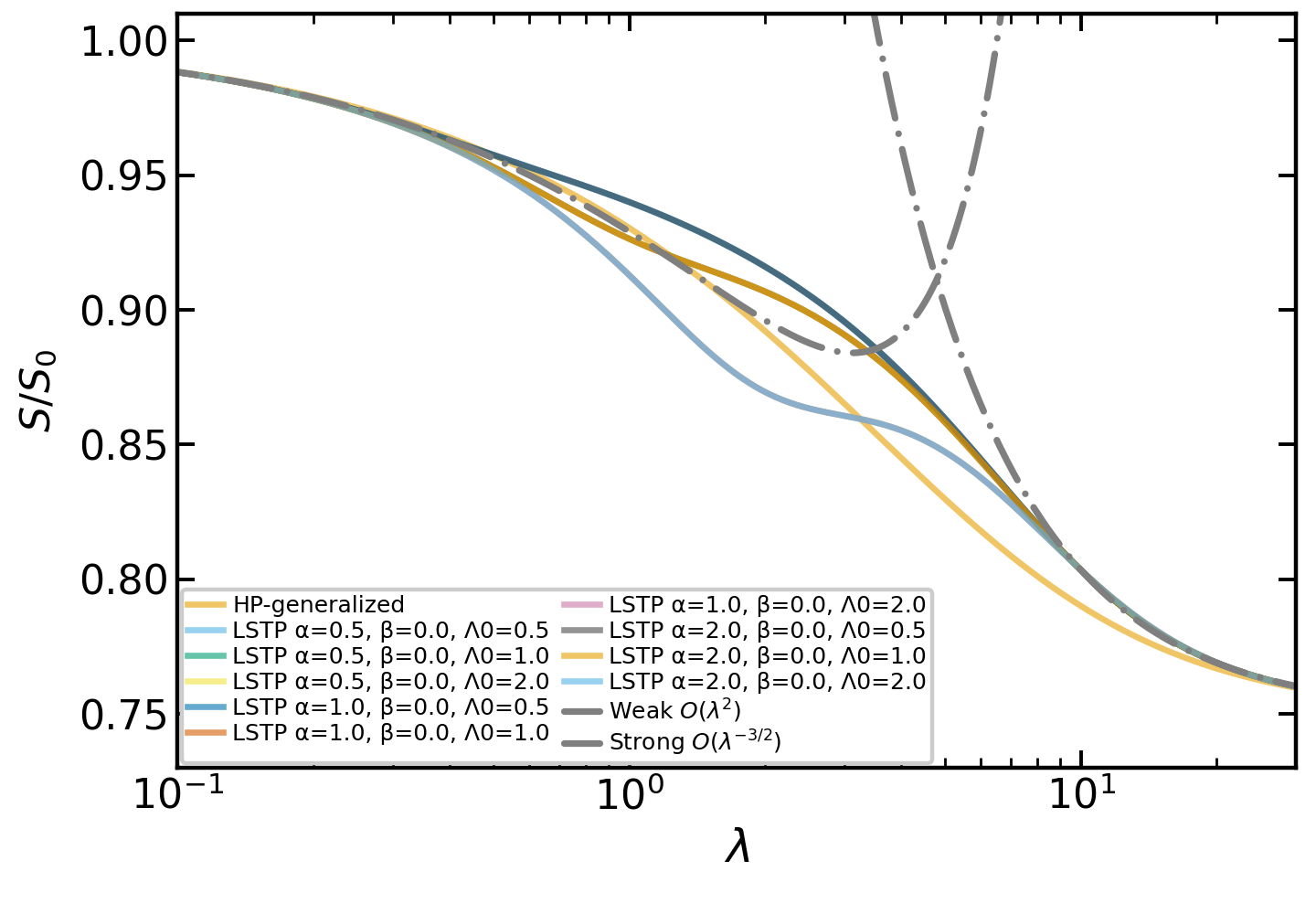}
  \end{minipage}

  \caption{All admissible individual curves (HP and LSTP) overlaid, together with the weak
  $\mathcal O(\Lam^{2})$ and strong $\mathcal O(\Lam^{-3/2})$ truncations.
  \textbf{Left:} full-range view.
  \textbf{Right:} zoomed view emphasizing the region where most of the variation occurs.
  The spread defines the admissible band.}
  \label{fig:allcurves}
\end{figure}

The full admissible band with the central curve is shown in Fig.~\ref{fig:band};
all individual survivors (HP and LSTP) are overlaid in Fig.~\ref{fig:allcurves}.
Figure~\ref{fig:allcurves} is a diagnostic of admissible interpolation freedom;
localized wiggles and double-peak structures in $d\mathcal S/d\ln\Lam$ for some
survivors are artifacts of the rational family and are not interpreted as physical.

\subsection{Predictions for unmeasured coefficients}
Beyond quantifying interpolation uncertainty, the ensemble \emph{predicts} higher-order series data on both sides.

\paragraph{\textbf{Weak-coupling $A_{5/2}$.}}
Let $f_{\mathrm{known}}(\Lam)$ denote the weak-coupling truncation in
Eq.~\eqref{eq:weak} through $\mathcal O(\Lam^{2})$, so that
\begin{equation}
\label{eq:f_remainder}
\qquad f(\Lam)=f_{\mathrm{known}}(\Lam)+A_{5/2}\,\frac{\Lam^{5/2}}{\pi^{5}}+\cdots\,.
\end{equation}
Because $e(\Lam)$ in Eq.~\eqref{eq:HP} inherits a logarithmic dependence
from $b(\Lam)$, the HP approximant generates a \emph{spurious}
$\Lam^{5/2}\Li$ contribution at the first unmatched order.
Expanding $f_{\mathrm{HP}}$ through $\mathcal O(\Lam^{5/2})$ gives
\begin{equation}
\label{eq:spur_log}
f_{\mathrm{HP}} = f_{\mathrm{known}}
 - \frac{3a}{\pi^{4}}\,\Lam^{5/2}\,\Li + r_{5,\mathrm{reg}}^{\mathrm{HP}}\,\Lam^{5/2} + \cdots,
\end{equation}
where $r_{5,\mathrm{reg}}^{\mathrm{HP}}$ denotes the nonlogarithmic $\Lam^{5/2}$
coefficient after separating the $\Lam^{5/2}\Li$ term, and $-3a/\pi^{4}\approx -0.0363$
is fully determined by the rational structure
($a$ is given in Eq.~\eqref{eq:HP_coeffs}).
The HP-generated $\Lam^{5/2}\Li$ term is an artifact of the
rational ansatz and should not be identified with the perturbative
weak-coupling structure.
A naive estimator $Q(\Lam)\equiv\pi^{5}\Lam^{-5/2}[f_{\mathrm{HP}}-f_{\mathrm{known}}]$
therefore diverges logarithmically as $\Lam\to 0$.
After subtracting the analytically known spurious log, the corrected estimator
\begin{equation}
\label{eq:Q_corr}
Q_{\mathrm{corr}}(\Lam)=\frac{\pi^{5}}{\Lam^{5/2}}\!
\left[f_{\mathrm{HP}}\!-\!f_{\mathrm{known}}+\frac{3a}{\pi^{4}}\,\Lam^{5/2}\,\Li\right]
\end{equation}
converges as $\Lam\to 0$, yielding
$A_{5/2}^{\mathrm{HP}}=\pi^{5}\,r_{5,\mathrm{reg}}^{\mathrm{HP}}$.
We evaluate $Q_{\mathrm{corr}}$ in extended-precision arithmetic at eleven points
$\Lam=10^{-3},10^{-4},\ldots,10^{-13}$ and extrapolate to $\Lam\to 0$
by fitting $Q_{\mathrm{corr}}=A_{5/2}^{\mathrm{HP}}+B\sqrt{\Lam}$
(the value is in principle obtainable in closed form from the rational
structure; the numerical evaluation serves as a convergence check on the
residual $\mathcal{O}(\sqrt{\Lam})$ remainder, which includes both
$\sqrt{\Lam}$ and $\sqrt{\Lam}\,\Li$ terms from the rational
structure; both vanish as $\Lam\to0$, and the fit-window stability
confirms they do not bias the intercept at the stated precision).
The fit is stable: restricting the window to
$\Lam\le10^{-4}$ or widening it to $\Lam=10^{-3}$ shifts the intercept
by less than $0.1$. We find
\begin{equation}
\label{eq:A52_pred}
\qquad \boxed{\;A_{5/2}^{\mathrm{HP}} = -43.8\pm 0.1\;}\,,
\end{equation}
where the uncertainty reflects fit-window variation.
We stress that $A_{5/2}^{\mathrm{HP}}$ is a prediction of the HP rational
ansatz and is therefore model-dependent; the true perturbative value of $A_{5/2}$
will be settled by an explicit $\mathcal O(\Lam^{5/2})$ EFT or
diagrammatic calculation.
The LSTP construction is not constrained beyond
$\mathcal O(\Lam^2\log\Lam)$ on the weak side and therefore does not yield a
reliable independent $A_{5/2}$.

\paragraph{\textbf{Finite-coupling estimator $\widehat S_{3}(\Lam_\ast)$ (admissibility bound).}}
At large $\Lam$,
\begin{equation}
\label{eq:strong_expansion}
\begin{aligned}
&\qquad f(\Lam)=\frac{3}{4}\Big[1+S_{3/2}\,\Lam^{-3/2}+S_{3}\,\Lam^{-3}+\mathcal O(\Lam^{-9/2})\Big],\\[4pt]
&\qquad S_{3/2}=\tfrac{15}{8}\,\zeta(3).
\end{aligned}
\end{equation}

Define the estimator
\begin{equation}
\label{eq:S3_estimator}
\begin{aligned}
&\qquad \widehat S_{3}(\Lam)=\Lam^{3}\!\Big[\tfrac{4}{3}\,f(\Lam)-1-S_{3/2}\,\Lam^{-3/2}\Big],\\
&\qquad S_{3}=\lim_{\Lam\to\infty}\widehat S_{3}(\Lam).
\end{aligned}
\end{equation}
so that $\widehat S_{3}(\Lam)=S_{3}+\mathcal O(\Lam^{-3/2})$.
Using only $f_{\min}\le f(\Lam)\le f_{\max}$ with $f_{\min}=0.75$ and $f_{\max}=1$, we obtain for any fixed $\Lam_\ast$,
\begin{equation}
\label{eq:S3_bound_interval}
\begin{aligned}
&\qquad \widehat S_{3}(\Lam_\ast)\in
\Big[\;\Lam_\ast^{3}\!\Big(\tfrac{4}{3}\,f_{\min}-1-S_{3/2}\,\Lam_\ast^{-3/2}\Big),\\
&\qquad\qquad\qquad \Lam_\ast^{3}\!\Big(\tfrac{4}{3}\,f_{\max}-1-S_{3/2}\,\Lam_\ast^{-3/2}\Big)\;\Big],
\end{aligned}
\end{equation}
Evaluated at $\Lam_\ast=10$, this yields
\begin{equation}
\label{eq:S3_bound}
\qquad\qquad\qquad\boxed{\,\widehat S_{3}(10)\in[-71.27,\,262.06]\,}.
\end{equation}

This interval is a direct consequence of the physical bounds
$0.75\le f\le1$, the known $S_{3/2}$, and the definition of the estimator, and
therefore provides a conservative, falsifiable benchmark for future holographic
computations at the reference point $\Lam_\ast=10$. The admissibility
bound in Eq.~\eqref{eq:S3_bound_interval} is formulated at a finite reference coupling
$\Lam_\ast$: it bounds the estimator $\widehat S_3(\Lam_\ast)$ using only these inputs.
With only this information, taking $\Lam_\ast$ larger does not strengthen the constraint.
In fact, Eq.~\eqref{eq:S3_bound_interval} implies a bound width
$\Delta \widehat S_3(\Lam_\ast)=\tfrac13 \Lam_\ast^3$, so the interval rapidly becomes
weaker as $\Lam_\ast$ increases (e.g.\ increasing $\Lam_\ast$ from 10 to 20 widens the
interval by a factor of 8). Taking $\Lam_\ast$ much smaller is also unhelpful, because one
then leaves the regime where the strong-coupling asymptotic expansion and the very notion
of an $S_3$ coefficient are under perturbative control. We therefore quote the interval at
$\Lam_\ast=10$ as a representative strong-coupling reference point where the constraint
remains nontrivial and our interpolants have already entered the asymptotic regime; other
choices follow from Eq.~\eqref{eq:S3_bound_interval} by the same scaling.

This also answers a 2021 private query on whether Pad\'e methods can anticipate the next
strong-coupling correction using existing information: applying the estimator in
Eq.~\eqref{eq:S3_estimator} together with the admissibility window $0.75\le f\le 1$ yields
Eq.~\eqref{eq:S3_bound}, i.e.\ $\widehat S_{3}(10)\in[-71.27,\,262.06]$. On the weak side,
the log-aware HP central extrapolation, after accounting for a spurious
$\Lam^{5/2}\log\Lam$ artifact (see below Eq.~\eqref{eq:spur_log}), yields
$A_{5/2}^{\mathrm{HP}}=-43.8\pm0.1$ in Eq.~\eqref{eq:A52_pred}. Together, these provide
concrete benchmarks to be tested and sharpened by future perturbative or holographic results.

\subsection{Summary of crossover scales}
Table~\ref{tab:crossovers} summarizes the curvature-defined crossover scale for
$\mathcal{S}/\mathcal{S}_0$.
Our entropy crossover $\Lam_c\simeq 3.52$ is consistent with Ref.~\cite{Muller:2025}
($\Lam_c=3.14$); the difference reflects the admissible-ensemble construction, which
shifts the inflection relative to a single-curve estimate.

\begin{table}[htb]
\caption{Crossover scale from curvature diagnostics.
$\Lam_\pm$ denote the admissible-ensemble bounds;
$F(\Lam_c)$ is the observable evaluated at $\Lam_c$.}
\label{tab:crossovers}
\begin{ruledtabular}
\begin{tabular}{lcccc}
Observable & $\Lam_c$ & $\Lam_-$ & $\Lam_+$ & $F(\Lam_c)$ \\
\hline
$\mathcal S/\mathcal S_0$ & 3.52 & 2.95 & 6.73 & 0.854 \\
\end{tabular}
\end{ruledtabular}
\end{table}

\section{Discussion}

We upgrade Pad\'e interpolation from a single-curve estimate to a \emph{controlled, admissible band}.
Both routes are explicitly \emph{log aware} and reproduce the \emph{full} weak-coupling expansion
through $\mathcal O(\Lam^2)$ \textit{exactly}---namely the coefficients of $\Lam$,
$\Lam^{3/2}$, $\Lam^2\log\Lam$, and the finite $\Lam^2$ term $A_{20}$.
On the strong side, the HP enforces $f\!\to\!3/4$, the exact vanishing of
$\Lam^{-1/2}$ and $\Lam^{-1}$ terms, and the known $\Lam^{-3/2}$ correction
$S_{3/2}=\tfrac{15}{8}\zeta(3)$.
The LSTP matching conditions enforce $f\!\to\!3/4$ and $S_{3/2}$, but the cutoff
tail introduces a controlled $\Lam^{-1}\Li$ artifact (Eq.~\eqref{eq:LSTP_leak}).
In addition, we exclude poles on the positive $\Lam$ axis, impose $0.75\!\le\! f\!\le\!1$ and
monotonicity in $\log\Lam$, and select the central curve by minimal curvature.
The Hermite-Pad\'e (HP) and log-subtracted two-point Pad\'e (LSTP) constructions use identical inputs but
different architectures. Their agreement, within a narrow band after admissibility filtering, is a
strong internal consistency check. Where they differ defines the admissible ensemble
uncertainty that any single-curve approach conceals.
 On the weak side we used the $\mathcal O(\Lam^2)$ series obtained via \emph{direct resummation}
in the Arnold-Zhai framework~\cite{Arnold:1994eb} (extended by us to $\mathcal N{=}4$ SYM) and then
rederived via EFT reconstruction in the Braaten-Nieto approach~\cite{Braaten:1995cm} (with our $\mathcal N{=}4$
implementation and corrected normalization of Ref.~\cite{Andersen:2021kld}). Relative to earlier Pad\'e analyses, we replace point estimates with a reproducible
admissible band. For entropy, the central crossover $\Lam_c\simeq 3.52$ agrees with previous single-curve
values, while we now quantify a realistic range, $\Lam_c\in[2.95,6.73]$, arising from admissible choices of
mapping and rational order.

Beyond uncertainty bands, the framework is \emph{predictive} without any new loop or higher-order
calculations. On the weak side, after analytically subtracting a spurious $\Lam^{5/2}\log\Lam$ artifact
of the HP rational structure (coefficient $-3a/\pi^{4}\approx-0.036$, see
Eq.~\eqref{eq:spur_log}), we infer $A_{5/2}^{\mathrm{HP}}=-43.8\pm0.1$ in the normalization
of Eq.~\eqref{eq:weak}. On the strong side, using only the known $S_{3/2}$ and the admissibility window
$0.75\le f\le1$, we obtain the model-independent interval
\[
\qquad \widehat S_{3}(10)\in[-71.27,\,262.06]\,,
\]
which directly addresses a 2021 query (private communication) on whether Pad\'e methods can anticipate
the next strong-coupling correction. The framework is modular: additional weak-side information
(e.g.\ $\mathcal O(\Lam^{5/2})$) or strong-side string corrections that fix $S_3$ (and beyond,
$\mathcal O(\Lam^{-9/2})$) will automatically shrink the band with no change in methodology.
\section{Outlook}
\label{sec:outlook}

A natural next step is to tighten and test the admissible ensemble band by incorporating higher-order terms in the weak-/strong-coupling expansions. Using EFT, the $\mathcal O(\Lam^{5/2})$ contribution to the free energy arises entirely from the soft scale $\sqrt{\Lam}\,T$ and is determined by three-loop vacuum diagrams in the electric effective theory, together with two-loop matching for the mass parameters $m_E^2$ and $m_S^2$ (cf.\ the QCD analysis in Ref.~\cite{Braaten:1995cm}). In parallel, one can extend the direct-resummation approach of Ref.~\cite{Du:2021jai}. Either route will fix $A_{5/2}$ and provide a sharp test of our HP prediction $A_{5/2}^{\mathrm{HP}}=-43.8\pm0.1$.

On the holographic side, the next unknown coefficient \(S_{3}\) arises from stringy
\(\alpha'\) corrections beyond the known \(\Lam^{-3/2}\) term. A computation of \(S_{3}\)
would turn our model-independent admissibility interval
\(\widehat S_{3}(10)\in[-71.27,\,262.06]\) into a definitive check of the
ensemble at strong coupling.

Extension to transport observables ($\eta/s$, $\hat q/T^3$, $2\pi T D_s$), where
admissible ensembles can systematically replace single-curve interpolations, is left to
future work.

After validation in \(\mathcal N{=}4\) SYM, the method can be applied to QCD, where the running
coupling and trace anomaly provide additional admissibility constraints, and to other gauge
theories with accessible weak and strong limits.
\begin{acknowledgments}
I am grateful to Juan Maldacena for insightful correspondence and for posing, in 2021, the question of whether Pad\'e interpolation could predict the next strong-coupling correction in thermal $\mathcal N{=}4$ SYM, which helped motivate part of this work. I also thank Michael Strickland and Qianqian Du, with whom I previously worked on  the single-curve Pad\'e interpolation for thermal $\mathcal N{=}4$ SYM. I also acknowledge support from the Department of Mathematics and Natural Sciences at Prince Mohammad Bin Fahd University.
\end{acknowledgments}

\appendix
\section{Analysis and reproducibility}\label{app:analysis}

\subsection{Numerical domain and grids}
We use $\Lam\in[10^{-4},10^{2}]$ on a uniform grid in $\log\Lam$ with at least 600 points.
Weak- and strong-coupling series are evaluated as in Sec.~\ref{sec:inputs}. All derivatives are
taken with respect to $\log\Lam$ using centered finite differences on the log grid.

\subsection{Route A (LSTP) admissible set}
For LSTP we take near diagonal $[m/n]=[4/4]$ and scan
\begin{equation}
\label{eq:STP_scan}
\begin{aligned}
&\qquad\alpha\in\{0.5,1.0,2.0\},\quad \beta\in\{0,0.05,0.1\},\\[3pt]
&\qquad\Lambda_0\in\{0.5,1.0,2.0,4.0\},\quad p=3.
\end{aligned}
\end{equation}
We subtract the weak-side logarithm with
$\chi(\Lam;\Lambda_0,p)=1/(1+(\Lam/\Lambda_0)^p)$ and approximate the residual by
$P_4(z)/Q_4(z)$ with $z$ from Eq.~\eqref{eq:map}. Coefficients are fixed by collocation at very
small and very large $\Lam$, then candidates are tested against the filters in Sec.~\ref{sec:filters}.
All admissible survivors have $\beta=0$, with $(\alpha,\Lambda_0)\in\{0.5,1.0,2.0\}\times\{0.5,1.0,2.0\}$
(\emph{none} at $\Lambda_0=4.0$); crossover data and pole margins are collected in
Tables~\ref{tab:summary} and~\ref{tab:polemargins}. We tested sensitivity to the conformal-map parameter by repeating the LSTP scan with
$\alpha\in\{1,2,4\}$ (instead of $\{0.5,1,2\}$), holding all other scan ranges fixed.
The admissible set and the resulting HP+LSTP envelope are unchanged within
$\mathcal{O}(10^{-6})$ over $\Lam\in[10^{-4},10^2]$, and the crossover diagnostics
are stable at the quoted precision. This invariance is expected
for $\beta=0$: since $z=y/(1+\alpha y)$ is a M\"obius
reparametrization, varying $\alpha$ does not alter the $[4/4]$
rational function class, and the matching conditions fix a unique
curve for each $\Lambda_0$.
At large $\Lam$, the HP-generalized curve lies slightly below the LSTP survivors.
This reflects the fact that the HP construction enforces the strong-coupling
structure globally in the mapped variable, whereas each LSTP interpolant is only
anchored to the strong expansion at a finite set of large-$\Lam$ matching points.
The resulting offset is small compared to the admissible-band width and decreases
rapidly with $\Lam$. The LSTP cutoff does introduce a residual
$\Lam^{-1}\Li$ artifact at large $\Lam$ (quantified in
Eq.~\eqref{eq:LSTP_leak}), but its prefactor is numerically negligible
within the admissible band on our working domain $\Lam\le10^{2}$.

\subsection{Route B (HP) central curve}
The HP generalized Pad\'e in Eq.~\eqref{eq:HP} matches the \emph{full} weak-coupling expansion
through $\mathcal O(\Lam^2)$ exactly (i.e., the coefficients of $\Lam$, $\Lam^{3/2}$,
$\Lam^2\log\Lam$, and the finite $\Lam^2$ term $A_{20}$). At strong coupling it reproduces
$f\!\to\!3/4$ and the $\Lam^{-3/2}$ correction $S_{3/2}=\tfrac{15}{8}\zeta(3)$, with the absence
of $\Lam^{-1/2}$ and $\Lam^{-1}$ enforced. The HP curve passes all admissibility checks and
minimizes the curvature functional in Eq.~\eqref{eq:central}; we therefore use it as the central
solution.
\subsection{Admissibility diagnostics}
\paragraph{\textbf{Bounds and monotonicity.}}
We require $0.75\le f(\Lam)\le 1$ and $\frac{d f}{d(\log\Lam)}\le 0$ on the interior window
$[10^{-3},10^{2}]$, while also checking the full domain for diagnostics.

\paragraph{\textbf{Pole exclusion.}}
We compute all roots of $Q_n(z)$ (and the HP denominator), map them to the $\Lam$ plane via
Eq.~\eqref{eq:map}, and exclude any poles on the positive real axis. Near cancelling Froissart
doublets are rejected. We report the minimal imaginary part among mapped poles (in the $\Lam$ plane)
as a safety margin.
\subsection{Crossover extraction}
We locate $\Lam_c$ by the log-space inflection condition, Eq.~\eqref{eq:lc}, using the zero of
$d^2 f/d(\log\Lam)^2$. For $\mathcal S/\mathcal S_0$ we
also quote an \emph{ensemble} crossover window using the pointwise envelope of the admissible set.

\begin{widetext}
    \subsection{Manual summary tables}
\end{widetext}

\begin{table}[htb]
  \caption{Admissible curves with crossover and value at crossover.}
  \label{tab:summary}
  \begin{ruledtabular}
  \begin{tabular}{lcccc}
  Curve & $\alpha$ & $\Lambda_0$ & $\Lam_c$ & $f(\Lam_c)$\\
  \hline
  HP-generalized & -- & -- & 3.52 & 0.854 \\
  \hline
  \multicolumn{5}{c}{\textit{LSTP survivors (all with $\beta=0$)}} \\
  LSTP & 0.5 & 0.5 & 6.45 & 0.839 \\
  LSTP & 0.5 & 1.0 & 6.73 & 0.834 \\
  LSTP & 0.5 & 2.0 & 2.95 & 0.861 \\
  LSTP & 1.0 & 0.5 & 6.45 & 0.839 \\
  LSTP & 1.0 & 1.0 & 6.73 & 0.834 \\
  LSTP & 1.0 & 2.0 & 2.95 & 0.861 \\
  LSTP & 2.0 & 0.5 & 6.45 & 0.839 \\
  LSTP & 2.0 & 1.0 & 6.73 & 0.834 \\
  LSTP & 2.0 & 2.0 & 2.95 & 0.861 \\
  \end{tabular}
  \end{ruledtabular}
  \end{table}
 
  \begin{table}[htb]
  \caption{Minimal imaginary part of mapped poles for LSTP survivors (all with $\beta=0$). Larger values indicate
  greater separation from the positive real $\Lam$ axis.}
 \label{tab:polemargins}
\begin{ruledtabular}
\begin{tabular}{ccc}
$\alpha$ & $\Lambda_0$ & Min.\ Im.\ part \\
\hline
0.5 & 0.5 & 7.38 \\
0.5 & 1.0 & 7.48 \\
0.5 & 2.0 & 8.24 \\
1.0 & 0.5 & 7.38 \\
1.0 & 1.0 & 7.48 \\
1.0 & 2.0 & 8.24 \\
2.0 & 0.5 & 7.38 \\
2.0 & 1.0 & 7.48 \\
2.0 & 2.0 & 8.24 \\
\end{tabular}
\end{ruledtabular}
\end{table}

\FloatBarrier


\begin{thebibliography}{99}

\bibitem{Du:2021jai}
Q.~Du, M.~Strickland, and U.~Tantary,
JHEP {\bf 08}, 064 (2021) [Erratum: JHEP {\bf 02}, 053 (2022)].

\bibitem{Andersen:2021kld}
J.~O.~Andersen, Q.~Du, M.~Strickland, and U.~Tantary,
Phys. Rev. D {\bf 105}, 015006 (2022).

\bibitem{Maldacena:1997re}
J.~M.~Maldacena,
Adv. Theor. Math. Phys. {\bf 2}, 231 (1998).

\bibitem{Gubser:1998bc}
S.~S.~Gubser, I.~R.~Klebanov, and A.~A.~Tseytlin,
Nucl. Phys. B {\bf 534}, 202 (1998).

\bibitem{Muller:2025}
B.~M\"uller,
Phys. Rev. D {\bf 112}, 054007 (2025).

\bibitem{BakerGravesMorris}
G.~A.~Baker, Jr. and P.~Graves-Morris,
\textit{Pad\'e Approximants}, 2nd ed.
(Cambridge University Press, Cambridge, England, 1996).

\bibitem{Chisholm1973}
J.~S.~R.~Chisholm,
Math. Comp. {\bf 27}, 841 (1973).

\bibitem{Guttmann1989}
A.~J.~Guttmann,
in \textit{Phase Transitions and Critical Phenomena}, Vol.~13,
edited by C.~Domb and J.~L.~Lebowitz
(Academic Press, New York, 1989), p.~1.

\bibitem{BoydSpectral}
J.~P.~Boyd,
\textit{Chebyshev and Fourier Spectral Methods}, 2nd ed.
(Dover, New York, 2001), chs.~7--9.

\bibitem{Arnold:1994eb}
P.~B.~Arnold and C.~X.~Zhai,
Phys. Rev. D {\bf 50}, 7603 (1994).

\bibitem{Braaten:1995cm}
E.~Braaten and A.~Nieto,
Phys. Rev. D {\bf 53}, 3421 (1996).

\end{thebibliography}
\end{document}